\documentclass[
journal=jacsat, 
manuscript=article]{achemso}

\usepackage[version=3]{mhchem} 

\usepackage{natbib}
\author{Mauricio Carrillo-Tripp}
\email{mauricio.carrillo@cinvestav.mx}
\affiliation[bmd]{Biomolecular Diversity Lab, Centro de Investigaci\'on y de Estudios Avanzados del Instituto Polit\'ecnico Nacional Unidad Monterrey, V\'ia del Conocimiento 201, Parque PIIT, C.P. 66600, Apodaca, Nuevo Le\'on, M\'exico}
\author{Leonardo Alvarez-Rivera}
\affiliation[itesi]{Ingenier\'ia en Sistemas, Instituto Tecnol\'ogico Superior de Irapuato, Guanajuato, M\'exico}
\author{Omar Israel Lara-Ram\'irez}
\affiliation[cc]{Departamento de Computaci\'on, Centro de Investigaci\'on y de Estudios Avanzados del Instituto Polit\'ecnico Nacional, Ciudad de M\'exico, M\'exico}
\author{Francisco Javier Becerra-Toledo}
\author{Adan Vega-Ram\'irez}
\author{Emmanuel Quijas-Valades}
\author{Eduardo Gonz\'alez-Zavala}
\author{Julio Cesar Gonz\'alez-V\'azquez}
\affiliation[itesi]{Ingenier\'ia en Sistemas, Instituto Tecnol\'ogico Superior de Irapuato, Guanajuato, M\'exico}
\author{Javier Garc\'ia-Vieyra}
\affiliation[cc]{Departamento de Computaci\'on, Centro de Investigaci\'on y de Estudios Avanzados del Instituto Polit\'ecnico Nacional, Ciudad de M\'exico, M\'exico}
\author{Nelly Beatriz Santoyo-Rivera}
\affiliation[itesi]{Ingenier\'ia en Sistemas, Instituto Tecnol\'ogico Superior de Irapuato, Guanajuato, M\'exico}
\author{Amilcar Meneses-Viveros}
\author{Sergio Victor Chapa-Vergara}
\affiliation[cc]{Departamento de Computaci\'on, Centro de Investigaci\'on y de Estudios Avanzados del Instituto Polit\'ecnico Nacional, Ciudad de M\'exico, M\'exico}

\title[HTMoL]{HTMoL: full-stack solution for remote access, visualization, and analysis of Molecular Dynamics trajectory data}

\begin{document}

\newpage

\begin{abstract}
The field of structural bioinformatics has seen significant advances with the use of Molecular Dynamics (MD) simulations of biological systems. The MD methodology has allowed to explain and discover molecular mechanisms in a wide range of natural processes. There is an impending need to readily share the ever-increasing amount of MD data, which has been hindered by the lack of specialized tools in the past. To solve this problem, we present HTMoL, a state-of-the-art plug-in-free hardware-accelerated web application specially designed to efficiently transfer and visualize raw MD trajectory files on a web browser. Now, individual research labs can publish MD data on the Internet, or use HTMoL to profoundly improve scientific reports by including supplemental MD data in a journal publication. HTMoL can also be used as a visualization interface to access MD trajectories generated on a high-performance computer center directly.\\
\textbf{Availability:} HTMoL is available free of charge for academic use. All major browsers are supported. A complete online documentation including instructions for download, installation, configuration, and examples is available at the HTMoL website http://htmol.tripplab.com. \\
\textbf{Supplementary information:} Supplementary data are available online. \\
\textbf{Corresponding author:} mauricio.carrillo@cinvestav.mx
\end{abstract}

\section{Introduction}

Fast acceleration in numerical processing speed has spread the use of computational techniques to study biological systems at the molecular level. It is expected that this tendency will grow with the advent of the so-called 'exascale' era. In particular, Molecular Dynamics (MD) has become a ubiquitous computational methodology in the life sciences, complementary to experimental techniques, because of its proved descriptive and predictive power. The use of MD is not restricted to study proteins \citep{COMPLEX_MD}, but also biological membranes \citep{MEM_MD}, genetic material \citep{RNA_MD}, drug design and discovery \citep{DRUGDESIGN_MD,DRUGDISCOV_MD}, and enhanced sampling strategies to study complex biological systems \citep{BIOLSYS_MD,MEMPROT_MD,STEERDRUGDISCOV_MD,SOTA_MD}.

Currently, there are several packages available to perform MD numerical simulations, e. g., Gromacs \citep{Gromacs}, Charmm \citep{Charmm}, Amber \citep{Amber}, or NAMD \citep{NAMD}. All of these packages are scalable and run in specialized high-performance parallel computing centers remotely. They generate large binary files containing the 3D space trajectory of all the atoms in the studied system as a function of time due to the intra- and inter-molecular forces. Statistical analysis of the trajectory file can be performed \emph{in situ}. However, the data has to be transferred to a local computer for visualization. Desktop applications are available for such purpose, e. g., VMD \citep{VMD} or  Pymol \citep{PyMOL}. Even though there are also some web applications for molecular visualization, e. g., JSMol \citep{JSmol} or NGL Viewer \citep{NGL}, which provide a wide range of features, they are only specialized for the exploration of static 3D structural data, usually in the form of PDB formatted files.

Sharing and publishing the raw MD trajectories of scientific research has been significantly hindered in the past. It is not trivial how to transfer the binary files from a server to the client and then visualize the three-dimensional time-dependent MD data directly on a web browser. Also, there has to be an efficient management of large amounts of data on client applications that can run on several kinds of computing devices, such as desktops, laptops, and mobile.

HTMoL solves these problems by the integration of state-of-the-art web technologies. HTMoL is simple to use, resulting in fast hardware-accelerated molecular graphics providing a clean graphical user interface (GUI) and a stable user experience (UX). 
Now, after installation on a web server and simple configuration through one file, HTMoL can be used by individual research labs to publish raw MD trajectory files on the Internet or embed the application on a web page with a single line of HTML code, features which are not found in any other molecular web visualizer.

\section{Methods}

\subsection{HTMoL Software architecture}

\begin{figure}[!tpb]
\centerline{\includegraphics[width=140mm]{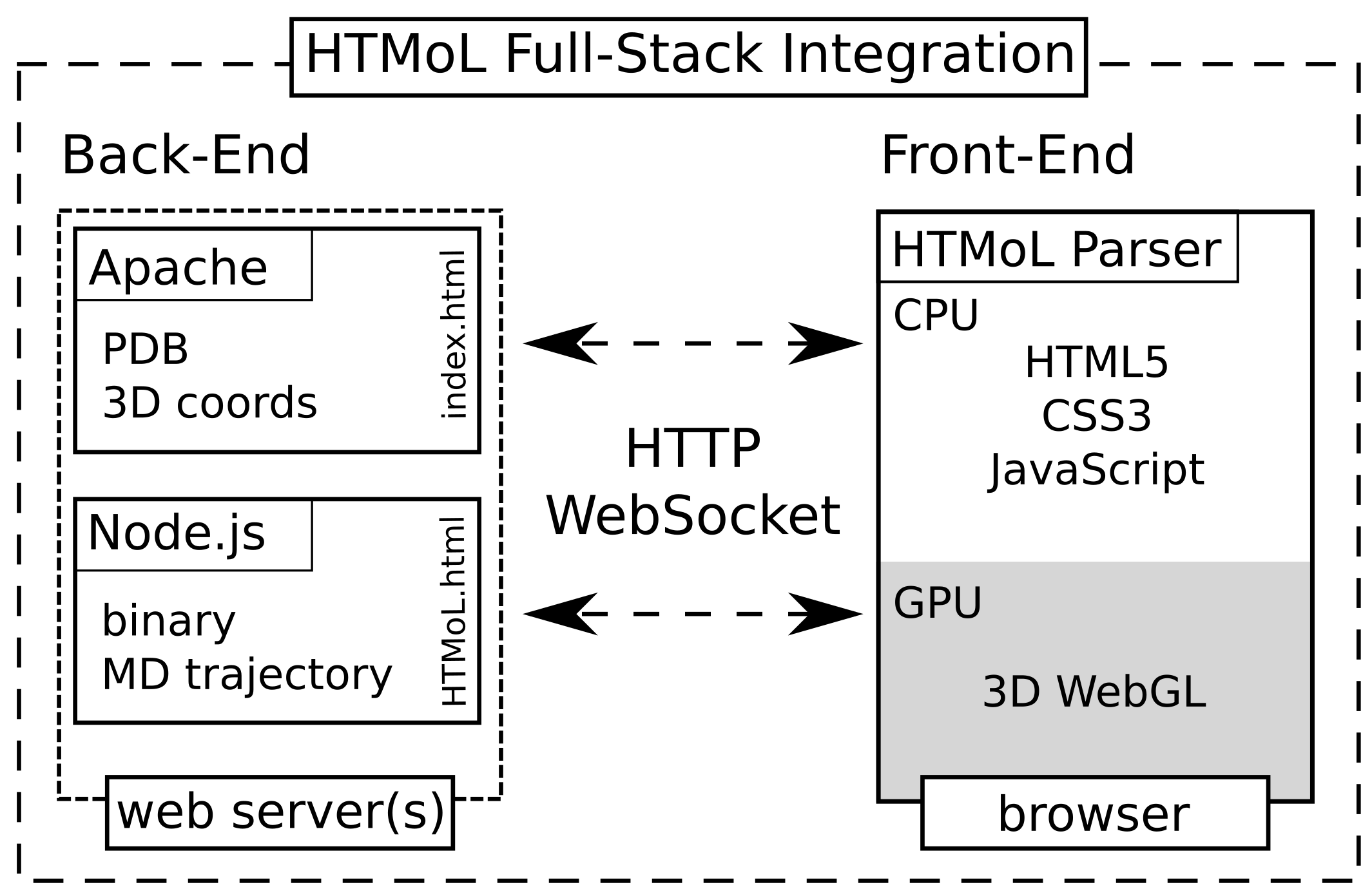}}
\caption{HTMoL software architecture. The molecular structure and dynamic data is transferred from the server(s) to the client through a Websocket connection. The data is then processed by the computer's GPU and displayed on a browser.}\label{S1}
\end{figure}

HTMoL is an efficient full-stack solution to the problems previously described by the integration of state-of-the-art web technologies (NodeJS+WebGL+JavaScript+HTML5+CSS3). Source code, documentation, and examples can be obtained at the HTMoL website. The term full-stack refers to an inclusion of all layers in computer software development. In other words, it involves code that runs in the web server (back-end) and code that runs in the client (front-end). An schematic of the global architecture of HTMoL is shown in Figure \ref{S1}.

On the back-end, it provides specialized code to send the raw MD trajectory file located on the server to the client over an Internet connection. On the front-end, HTMoL provides methods to parse and interpret molecular dynamics data on the client, using hardware-accelerated computational techniques to interactively display the molecular three-dimensional structure and dynamics on a web browser through a GUI, achieving high performance. MD data is processed in the input stream to avoid local storage on the client side. 


\subsection{Back-end}

In HTMoL, the back-end follows a microservice pattern. The initial PDB formatted atomic coordinates, defined on an ASCII text file, are handled by the Apache server over a standard HTTP connection, as usually done. However, the MD trajectory files have to be managed by a Node.js server because it provides the methods to transfer binary data highly efficiently through a WebSocket with the help of the minimalist web frameworks express.js and binary.js. Node.js is a JavaScript runtime built on Chrome's V8 engine. It uses an asynchronous event-driven, non-blocking I/O model that makes it lightweight. WebSockets are a way for the client to communicate back and forth with the web server without all the overhead of a standard HTTP connection. Even though it might be more convenient in some cases, the Apache and Node.js servers do not have to be running on the same machine.

\subsection{Front-end}

The HTMoL functionality is written in JavaScript. The GUI is built with HTML5 and CSS3. A service worker handles the WebSocket connection with the binary server. Once received, the visualization of the molecular structure and dynamics data is handled by the hardware-accelerated WebGL rasterization engine. Since WebGL runs on the graphics processing unit (GPU) of the client's computer, HTMoL provides the necessary code in the form of two primary functions called the vertex shader and the fragment shader. The first one processes the vertex positions that can then rasterize various kinds of primitives, i. e., points, lines, and triangles. In turn, a set of primitives defines a surface mesh of a complex object, e. g., a sphere or cylinder. The second function computes the color for each pixel of the primitive being drawn. After setting up the atom's positions and representations through the WebGL API, the two functions are executed by a call to gl.drawArrays, which computes the shaders on the GPU. All the molecule's structural data is handed over to the GPU through attributes and buffers. Every frame found in the trajectory file by the worker is then processed through a loop, updating the atoms positions on screen at every MD step. This integration results in a robust UX which allows performing several visualization options in real time.

\subsection{Implementation and features}

\begin{figure}[!tpb]
\centerline{\includegraphics[width=160mm]{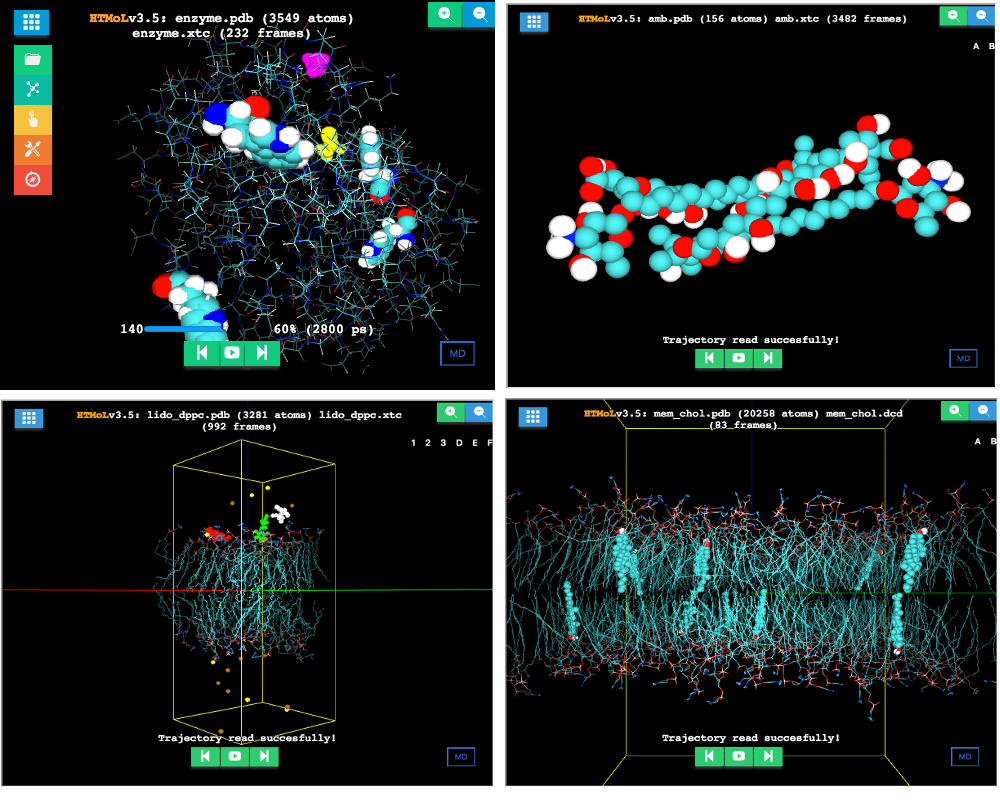}}
\caption{HTMoL GUI, displaying the MD trajectory of previously reported works (clockwise): an enzyme (lines representation), with N-, C- terminals, and residues of type THR highlighted (VDW representation using magenta, yellow, and atom name colors, respectively) \citep{MCT2013}; an Amphotericin B drug dimer inter-molecular interaction (VDW representation, atom name colors) \citep{MCT2016}; a biological membrane (phospholipids in line representation, cholesterol molecules in VDW representation, atom name colors) \citep{MCT2005}; three anesthetic drug molecules (VDW represetnation, chain name colors) interacting with a DPPC bilayer (line representation, atom name colors), where the Cartesian axes and computational cell box are shown (in press). In all cases, the water molecules were removed from the trajectory file for clarity. These examples are accessible through the HTMoL website}\label{S2}
\end{figure}

Since the HTMoL functionality is written in JavaScript, there is no need to compile. The server administrator (back-end-user) only has to edit a configuration file (local/config.js) to set the values of parameters that control the network (providing a public IP and an open Port) and the MD data location (providing the path to the PDB and trajectory files). Once the Node.js is started by running the provided script (BinServer.js), the MD data is readily accessible on the corresponding server's URL through a web browser. Some examples are shown in Figure \ref{S2}.

HTMoL can be used through the HTMoL.html file as is, or embedded on a webpage with the addition of a single $<iframe>$ HTML tag in the body of the index.html file. HTMoL can also be configured to extend the File menu allowing the front-end-user to specify the name of the PDB and trajectory files to be loaded, and an option to enable downloading the data to the client's computer. This configuration flexibility is essential if HTMoL is to be used as a visualization interface of a high-performance computer cluster were the MD data is generated.

\begin{figure}[!tpb]
\centerline{\includegraphics[width=115mm]{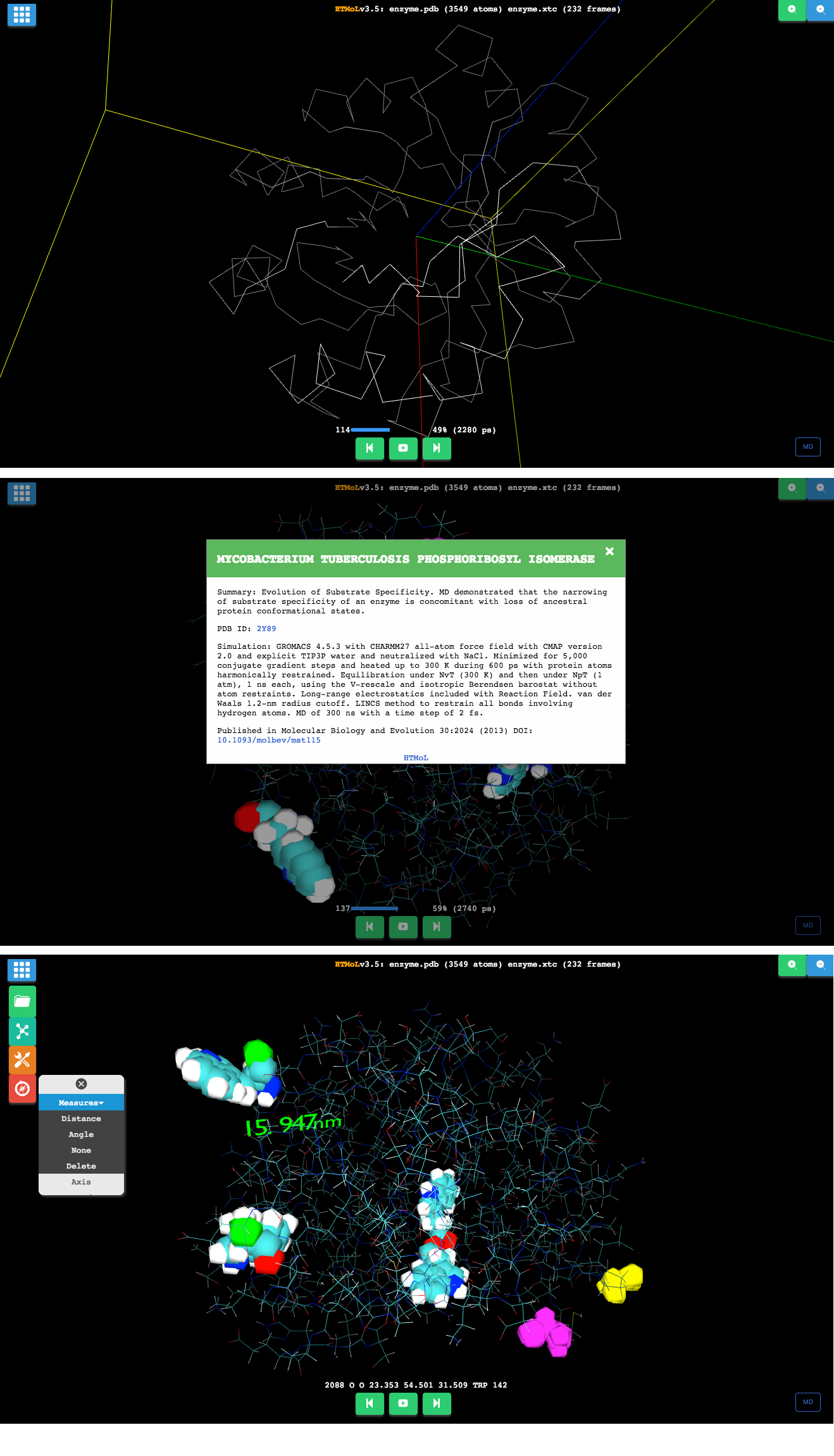}}
\caption{HTMoL full-size GUI (from top to bottom): protein's backbone trace representation, showing the computational box (yellow) and the Cartesian coordinate axes (X in red, Y in green, and Z in blue); MD simulation details; inter-atomic distance measurement tool, located in the Actions menu. Since these features run independently of the trajectory controls, the front-end user can access them while the MD is played in the background.}\label{S3}
\end{figure}

Also, HTMoL implements a nondeterministic finite automaton (NFA) to make atom selections by a set of commands parsed through a syntactic analyzer. The selection can then be modified by a variety of visualization options. These include different geometries (e. g. line, CPK, VDW and back-bone trace), different colorings (name or RGB values), and various levels of abstraction (control of the trajectory, hiding or showing chains, zooming in or out, different points of view). The simulation's computational cell box can be shown if its dimensions are defined in the PDB file. It is possible to provide information on the molecular system and the MD simulation details. These options are defined in the back-end's configuration file. HTMoL's front-end GUI offers a graphical menu and buttons to interact with the application. Clicking on an atom will show related information. Measurement of distance and angles between atoms is also possible via the Action menu (Figure \ref{S3}).

\section{Results}

The current HTMoL function and feature set achieves its primary goal; access and visualize MD data stored on a remote server through a web browser. In following versions, HTMoL will provide more representation types for bonds and atoms (e. g. licorice, hyperballs), secondary structure depictions based on a protein's backbone atoms (e. g. cartoon, tube, ribbon, rocket), as well as methods to directly support coarse-grained models.

To the best of our knowledge, HTMoL is one of its kind. For this reason, a comparison with other applications is not possible at this time. However, internal tests showed that the use of third-party libraries like \emph{three.js} to create and display animated 3D computer graphics on the web browser, although easier to implement, have a detrimental impact on the application's performance. In its latest version (3.5), HTMoL abandoned the use of such frameworks. Instead, now it has code with native primitives built-in, i. e., atoms and bonds in different representations. This strategy consistently increased the rasterization performance one order of magnitude in all browsers and platforms tested (FireFox v57, Chrome v62, Safari v9, in Linux Ubuntu, Mac OS, and Windows).

HTMoL is under constant development and improvement. Currently, the MD trajectory of a system of 20,000 atoms runs at $>$30 FPS on FireFox Quantum in a MacBook laptop (50\% of optimal performance). The performance will increase with the implementation of the ray-casted impostors strategy which reduces the geometric complexity of the system. This, together with the constant increase of web browsers and GPU speed, will allow the trajectory display of very large systems ($10^5 - 10^6$ atoms). We also expect the efficiency to increase with the future implementation of new technologies like webassembly and webpack, or techniques like HTML web components or progressive application development to run on mobile devices providing a unique UX.

\section{Conclusion}

HTMoL can be considered to be the first of a next-generation Molecular Dynamics visualization web applications. It pioneers on the leverage of state-of-the-art web technologies to produce an efficient platform to transfer and visualize raw MD data on a browser with good performance. It sets the solid foundations of an evolving full-stack application. HTMoL has the potential to change the MD data sharing paradigm in the field of structural bioinformatics due to its simple yet powerful architecture.


\section*{Funding}

This work has been supported by the Consejo Nacional de Ciencia y Tecnolog\'ia M\'exico (grant number 132376 to M.C.-T.).








\newpage

\newpage
\clearpage

\bibliography{HTMoLv3.5}

\end{document}